\documentclass{ws-jnmp}
\usepackage{amsfonts}

\begin{document}

\newcommand{\myfrac}[2]{\frac{\displaystyle #1}{\displaystyle #2}}
\newcommand{\myA}{A}
\newcommand{\myB}{B}
\newcommand{\myC}{C}
\newcommand{\mytildeB}{\widetilde\myB}
\newcommand{\myS}{\mathbb{S}}
\newcommand{\mymatrix}[1]{\begin{pmatrix}#1\end{pmatrix}}
\renewcommand{\Re}{\mathop{\mbox{Re}}}
\renewcommand{\Im}{\mathop{\mbox{Im}}}
\renewcommand{\vec}[1]{\boldsymbol{#1}}

\markboth{G.M. Pritula and V.E. Vekslerchik}
{Toda-Heisenberg chain}

%
\catchline{}{}{2010}{}{}
%

\copyrightauthor{G.M. Pritula and V.E. Vekslerchik}

\title{Toda-Heisenberg chain: interacting $\sigma$-fields in two dimensions.}

\author{G.M. Pritula$^\dag$ and V.E. Vekslerchik$^\ddag$}

\address{
Usikov Institute of Radiophysics and Electronics \\
12, Proskura st., Kharkov, 61085, Ukraine \\
\email{$^\dag$galinapritula@yandex.ru \\ $^\ddag$vekslerchik@yahoo.com}
}

\maketitle

\begin{history}
\received{}
\revised{}
\accepted{}
\end{history}

\begin{abstract}
We study a (2+1)-dimensional system that can be viewed as an infinite number 
of $O(3)$ $\sigma$-fields coupled by a nearest-neighbour Heisenberg-like 
interaction. We reduce the field equations of this model to an integrable 
system that is closely related to the two-dimensional relativistic Toda chain 
and the Ablowitz-Ladik equations. Using this reduction we obtain the 
dark-soliton solutions of our model.
\end{abstract}


\ccode{2010 Mathematics Subject Classification:  37J35, 35Q51, 37K10,  37K35, 11C20}

\makeatletter \@addtoreset{equation}{section} \makeatother

\section{Introduction.}

The model considered in this paper can be viewed as a generalization of the 
classical $O(3)$ $\sigma$-model in two dimensions, described by the Hamiltonian 
function
\begin{equation}
  \mathcal{E} = 
  \mathcal{E}[\vec{\sigma}] = 
  \int\limits_{\mathbb{R}^{2}} dx \, dy \;
  \left( \nabla\vec{\sigma}, \nabla\vec{\sigma} \right)
\end{equation}
where $\vec{\sigma}$ is a three-component vector of unit length,
\begin{equation}
  \left( \vec{\sigma}, \vec{\sigma} \right) = 1.
\end{equation}
and braces denote the standard scalar product.
The energy of our system is given by 
\begin{equation}
  \mathcal{H} = 
  \sum_{n}\mathcal{E}\left[\vec{\sigma}_{n}\right] 
  +
  \mathcal{H}_{int} 
\end{equation}
with nearest-neighbour interaction 
\begin{equation}
  \mathcal{H}_{int} = 
  \frac{1}{2}
  \sum_{n}\sum_{p = n \pm 1} \mathcal{U}_{np} 
\end{equation}
of the Heisenberg type:
\begin{equation}
  \mathcal{U}_{np}
  = 
  \mathcal{U}\left[ \vec{\sigma}_{n}, \vec{\sigma}_{p} \right] 
  =
  \int\limits_{\mathbb{R}^{2}} dx \, dy \;
  F\biggl(\left( \vec{\sigma}_{n}, \vec{\sigma}_{p} \right)\biggr). 
\end{equation}
The models of this type can appear, for example, in the studies of the lamellar
(graphite-like) magnetics when the spin interaction inside one layer can be described 
in the framework of the Landau-Lifshitz theory with effective Heisenberg interaction 
between adjacent layers. 

The stationary structures of our system are governed by the (2+1)-dimensional 
equation
\begin{equation}
  \frac{ \delta \mathcal{H} }{ \delta \vec{\sigma}_{n} } = 0,
  \qquad
  \left(\delta\vec{\sigma}_{n}, \vec{\sigma}_{n} \right) = 0. 
\end{equation}
In what follows we use a function $F$ which is peculiar to integrable 
nonlinear mathematics (see \textit{e.g.} \cite{IK1981,S1982}), 
\begin{equation}
  F(x) = g^{2}\ln( 1 + x).
\end{equation}
The resulting equations are given by
\begin{equation}
  \left[ \Delta \vec\sigma_{n}, \vec\sigma_{n} \right] =
  \frac{g^{2}}{4}  
  \sum_{p = n \pm 1} f_{np} 
  \left[ \vec\sigma_{p}, \vec\sigma_{n} \right] 
\end{equation}
where
\begin{equation}
  f_{np} = \frac{ 2 }{ 1 + \left( \vec\sigma_{n}, \vec\sigma_{p} \right) }. 
\label{def-f-nn}
\end{equation}
The factor $g^{2}$ can be eliminated by rescaling the coordinates, so we take   
\begin{equation}
  g = 4
\end{equation}
and write the central equation of our study as
\begin{equation}
  {\scriptstyle \frac{1}{4}} 
  \left[ \Delta \vec\sigma_{n}, \vec\sigma_{n} \right] =
  \sum_{p = n \pm 1} f_{np} 
  \left[ \vec\sigma_{p}, \vec\sigma_{n} \right]. 
\label{eq-htc-vector}
\end{equation}

In the following sections, after re-parametrization of (\ref{eq-htc-vector}), 
we split it in Sec. \ref{sec-splitting} into a first-order system, bilinearize it 
(Sec. \ref{sec-bilin}) and derive the dark-soliton solutions 
(Sec. \ref{sec-solitons}).

\section{Parametrization and splitting \label{sec-splitting}}

Using the vector-matrix correspondence
\begin{equation}
  \vec{\sigma} = 
  \left( s_{1}, s_{2}, s_{3} \right)^{T} 
  \qquad \to \qquad
  \myS = \mymatrix{ s_{3} & s_{1} - i s_{2} \cr s_{1} + i s_{2} & - s_{3} }
  = \sum_{j=1}^{3} s_{j}\sigma^{j}
\label{vmc}
\end{equation}
where $\sigma^{j}$ ($j=1,2,3$) are the Pauli matrices 
\begin{equation}
  \sigma^{1} = \mymatrix{ 0 & 1 \cr 1 & 0 },
  \qquad
  \sigma^{2} = \mymatrix{ 0 & -i \cr i & 0 },
  \qquad
  \sigma^{3} = \mymatrix{ 1 & 0 \cr 0 & -1 } 
\end{equation}
and introducing complex variables
\begin{equation}
  z = x + iy,
  \qquad
  \bar{z} = x - iy
\end{equation}
one can rewrite Eq. (\ref{eq-htc-vector}) as 
\begin{equation}
  \left[ \partial\bar\partial \, \myS_{n}, \myS_{n} \right]
  = 
  \sum_{p = n \pm 1} 
    f_{np}
    \left[ \myS_{p}, \myS_{n} \right]
\label{eq-main-matrix}
\end{equation}
with 
$\partial = \partial/\partial z$, 
$\bar\partial = \partial/\partial \bar{z}$ 
and
\begin{equation}
  f_{np} = 
  \frac{ 2 }
       { 1 + {\scriptstyle\frac{1}{2} } \mathop{\mbox{tr}} \myS_{n}\myS_{p} }.
\end{equation}
In what follows we use the parametrization of the vectors $\vec{\sigma}_{n}$ 
based on the presentation of the matrices $\myS_{n}$ in the form
\begin{equation}
  \myS_{n} = \Psi_{n}^{-1} \sigma^{3} \, \Psi_{n}.
\end{equation}
Using the invariance of this representation with respect to transformations 
$\Psi_{n} \to \mathbb{D}_{n} \Psi_{n}$ with arbitrary diagonal matrices 
$\mathbb{D}_{n}$ one can choose
\begin{equation}
  \Psi_{n} = \mymatrix{ 1 & \myB_{n} \cr \myC_{n} & 1 } 
\end{equation}
which leads to
\begin{equation}
  \myS_{n} 
  = 
  \frac{ 1}{ 1 - \myB_{n}\myC_{n} }
  \mymatrix{ 1 + \myB_{n}\myC_{n} & 2\myB_{n} \cr 
  -2\myC_{n} & - 1 - \myB_{n}\myC_{n} }. 
\end{equation}
Calculating $\partial\bar\partial \, \myS_{n}$ and $f_{np}$,
\begin{equation}
  f_{np} = 
  \frac{\left(1 - \myB_{n}\myC_{n}\right)\left(1 - \myB_{p}\myC_{p} \right)}
       {\left(1 - \myB_{n}\myC_{p}\right)\left(1 - \myC_{n}\myB_{p} \right)},
\end{equation}
one comes to the following system of equations:
\begin{equation}
  \left\{
  \begin{array}{lcl}
  \myA_{n} \mathcal{L}^{\myB}_{n} & = & 
  \bar{Y}_{n} \left( \myB_{n+1} - \myB_{n} \right)
  -
  Y_{n-1} \left( \myB_{n} - \myB_{n-1} \right)
  \\[2mm]
  \myA_{n} \mathcal{L}^{\myC}_{n} & = & 
  Y_{n} \left( \myC_{n+1} - \myC_{n} \right)
  -
  \bar{Y}_{n-1} \left( \myC_{n} - \myC_{n-1} \right)
  \end{array}
  \right.
\label{syst-BC}
\end{equation}
where 
\begin{equation}
  \myA_{n} = \frac{ 1 }{ 1 - \myB_{n}\myC_{n} },
\end{equation}
\begin{eqnarray} 
  \mathcal{L}^{\myB}_{n} & = & 
  \partial \bar\partial \myB_{n} + 
  2 \myA_{n} 
    \left( \partial \myB_{n} \right) 
    \left( \bar\partial \myB_{n} \right) 
    \myC_{n}, 
  \\[2mm]
  \mathcal{L}^{\myC}_{n} & = & 
  \partial \bar\partial \myC_{n} + 
  2 \myA_{n} 
    \myB_{n} 
    \left( \partial \myC_{n} \right) 
    \left( \bar\partial \myC_{n} \right) 
\end{eqnarray}
and
\begin{equation}
  Y_{n} = \frac{ 1 }{ 1 - \myB_{n} \myC_{n+1} }, 
  \qquad 
  \bar{Y}_{n} = \frac{ 1 }{ 1 - \myB_{n+1} \myC_{n} }. 
\end{equation}

The crucial step of our proceeding is the following \textit{ansatz}:
we split the above system into two first-order ones,
\begin{equation}
  \left\{
  \begin{array}{lcl}
  i \myA_{n} \partial \, \myB_{n} & = & 
  Z_{n-1} \left( \myB_{n} - \myB_{n-1} \right) 
  \\
  i \myA_{n} \partial \, \myC_{n} & = & 
  Z_{n} \left( \myC_{n+1} - \myC_{n} \right) 
  \end{array} 
  \right.
\label{system-ansatz-P}
\end{equation}
and
\begin{equation}
  \left\{
  \begin{array}{lcl}
  -i \myA_{n} \bar\partial \, \myB_{n} & = & 
  \bar{Z}_{n} \left( \myB_{n+1} - \myB_{n} \right) 
  \\
  -i \myA_{n} \bar\partial \, \myC_{n} & = & 
  \bar{Z}_{n-1} \left( \myC_{n} - \myC_{n-1} \right). 
  \end{array} 
  \right.
\label{system-ansatz-N}
\end{equation}
By direct calculations one can show that this can be done provided we can find 
the functions $Z_{n}$ and $\bar{Z}_{n}$ that (i) make 
(\ref{system-ansatz-P}) and (\ref{system-ansatz-N}) compatible and 
(ii) lead to (\ref{syst-BC}). 
It is demonstrated in the appendix that the functions $Z_{n}$ and $\bar{Z}_{n}$ 
that meet these conditions can be chosen as 
\begin{equation}
  Z_{n} = \zeta \, Y_{n}, 
  \qquad
  \bar{Z}_{n} = \bar\zeta \, \bar{Y}_{n}  
\end{equation}
where $\zeta$ and $\bar\zeta$ are arbitrary constants related by 
\begin{equation}
  \zeta \bar\zeta = 1 
\end{equation}
To summarize, one can obtain a large number of solutions of (\ref{eq-main-matrix})
by solving the system 
\begin{equation}
  \left\{
  \begin{array}{lcl}
  i \partial \myB_{n} & = & 
  \zeta \; 
  \displaystyle\frac{ 1 - \myB_{n}\myC_{n} }{ 1 - \myB_{n-1}\myC_{n} } 
  \left( \myB_{n} - \myB_{n-1} \right) 
  \\[4mm]
  i \partial \myC_{n} & = & 
  \zeta \; 
  \displaystyle\frac{ 1 - \myB_{n}\myC_{n} }{ 1 - \myB_{n}\myC_{n+1} } 
  \left( \myC_{n+1} - \myC_{n} \right) 
  \end{array} 
  \right.
\label{system-P}
\end{equation}
and
\begin{equation}
  \left\{
  \begin{array}{lcll}
  -i \bar\partial \myB_{n} & = & 
  \bar\zeta \; 
  \displaystyle\frac{ 1 - \myB_{n}\myC_{n} }{ 1 - \myB_{n+1}\myC_{n} } 
  \left( \myB_{n+1} - \myB_{n} \right) 
  \\[4mm]
  -i \bar\partial \myC_{n} & = & 
  \bar\zeta \; 
  \displaystyle\frac{ 1 - \myB_{n}\myC_{n} }{ 1 - \myB_{n}\myC_{n-1} } 
  \left( \myC_{n} - \myC_{n-1} \right) 
  \end{array} 
  \right.
\label{system-N}
\end{equation}

Before proceed further, we would like to give some comments on this system.
After introducing new variables, 
\begin{equation}
  \mytildeB_{n} = 1 / \myC_{n},
\end{equation}
Eqs. (\ref{system-P}), (\ref{system-N}) can be cast into the Hamiltonian form
\begin{equation}
  \left\{
  \begin{array}{lcl}
  i\partial \myB_{n} & = & 
  \left( \myB_{n} - \mytildeB_{n} \right)^{2} 
  \displaystyle\frac{ \partial H }{ \partial \mytildeB_{n} }
  \\[4mm]
  - i\partial \mytildeB_{n} & = & 
  \left( \myB_{n} - \mytildeB_{n} \right)^{2} 
  \displaystyle\frac{ \partial H }{ \partial \myB_{n} }
  \end{array}
  \right.
\end{equation}
with
\begin{equation}
  H = 
  \zeta 
  \sum_{n=-\infty}^{\infty} 
    \ln\frac{ \myB_{n} - \mytildeB_{n} }{ \myB_{n} - \mytildeB_{n+1} }
\end{equation}
and
\begin{equation}
  \left\{
  \begin{array}{lcl}
  i\bar\partial \myB_{n} & = & 
  \left( \myB_{n} - \mytildeB_{n} \right)^{2} 
  \displaystyle\frac{ \partial \bar{H} }{ \partial \mytildeB_{n} }
  \\[4mm]
  - i\bar\partial \mytildeB_{n} & = & 
  \left( \myB_{n} - \mytildeB_{n} \right)^{2} 
  \displaystyle\frac{ \partial \bar{H} }{ \partial \myB_{n} }
  \end{array}
  \right.
\end{equation}
with
\begin{equation}
  \bar{H} = 
  \bar\zeta 
  \sum_{n=-\infty}^{\infty} 
    \ln\frac{ \myB_{n} - \mytildeB_{n} }{ \myB_{n} - \mytildeB_{n-1} }
\end{equation}
and can be identified with the $\left( X_{1}, Y_{1} \right)$ equations 
(with $a(u,v) = (u-v)^{2}$) from the list of the paper by Adler and Shabat 
\cite{AS2006}.

At the same time both $\myB_{n}$ and $\myC_{n}$ solve the (2+1)-dimensional 
version of the Ruijsenaars-Toda lattice \cite{F1997,SY1997}
\begin{equation}
  \partial\bar\partial U_{n} 
  + 
  \left( \partial U_{n} \right)
  \left( \bar\partial U_{n} \right)
  \left[
    \frac{ 1 }{ U_{n+1} - U_{n} } 
    -
    \frac{ 1 }{ U_{n} - U_{n-1} } 
  \right]
  = 0. 
\label{ruij-toda-lattice}
\end{equation}
Note that equations (\ref{ruij-toda-lattice}) are different from 
(and complementary to) the Ruijsenaars-Toda lattice $(R_{1})$ that appears in 
a natural way in the framework of \cite{AS2006}.

Finally, calculating from Eqs. (\ref{system-ansatz-P}), (\ref{system-ansatz-N}) 
derivatives of the functions $f_{n}$ defined by  
\begin{equation}
  f_{n} = f_{n,n+1} =
  \frac{ 
    \left( 1 - \myB_{n}\myC_{n} \right) 
    \left( 1 - \myB_{n+1}\myC_{n+1} \right) 
  }{
    \left( 1 - \myB_{n}\myC_{n+1} \right) 
    \left( 1 - \myB_{n+1}\myC_{n} \right) 
  } 
\end{equation}
one can demonstrate that these functions satisfy 
\begin{equation}
  \partial\bar\partial \ln f_{n} = 
  f_{n+1} - 2 f_{n} + f_{n-1}. 
\end{equation}
Thus one can see the relationship of the model discussed in this paper with 
the famous two-dimensional Toda lattice.

\section{Bilinearization. \label{sec-bilin}}

\newcommand{\mybetaB}{\beta}
\newcommand{\mybetaC}{\beta}
\newcommand{\mygammaA}{\gamma^{\myA}}
\newcommand{\mygammaB}{\gamma^{\myB}}
\newcommand{\mygammaC}{\gamma^{\myC}}

To bilinearize Eqs. (\ref{system-P}), (\ref{system-N}) we introduce 
$\check\rho_{n}$, $\check\tau_{n}$, $\hat\tau_{n}$ and $\hat\sigma_{n}$
by
\begin{equation}
  \myB_{n} = \frac{ \check\rho_{n-1} }{ \hat\tau_{n} },
  \qquad 
  \myC_{n} = - \frac{ \hat\sigma_{n} }{ \check\tau_{n-1} }
\end{equation}
and another set of tau-functions by
\begin{equation}
  \left\{
  \begin{array}{lcl}
  i D \, \check\rho_{n-1} \cdot \hat\tau_{n} & = & 
  \alpha \; \rho_{n-1}\tau_{n} 
  \\
  i D \, \check\tau_{n-1} \cdot \hat\sigma_{n} & = & 
  \alpha \; \tau_{n-1}\sigma_{n} 
  \end{array} 
  \right.
\label{eq-tau-alpha}
\end{equation}
and 
\begin{equation}
  \left\{
  \begin{array}{lcl}
  -i \bar{D} \, \check\rho_{n-1} \cdot \hat\tau_{n} & = & 
  \bar\alpha \; \tau_{n-1}\rho_{n} 
  \\
  - i \bar{D} \, \check\tau_{n-1} \cdot \hat\sigma_{n} & = & 
  \bar\alpha \; \sigma_{n-1}\tau_{n} 
  \end{array} 
  \right.
\label{eq-tau-bar-alpha}
\end{equation}
where $\alpha$ and $\bar\alpha$ are constants, 
$D$ and $\bar{D}$ are the Hirota's bilinear differential operators,
$
  D \, u  \cdot v = 
  \left( \partial u \right) v - u \left( \partial v \right) 
$
and 
$
  \bar{D} \, u  \cdot v = 
  \left( \bar\partial u \right) v - u \left( \bar\partial v \right). 
$
Now, to finish the bilinearization of our equations, we impose the restrictions
\begin{equation}
  \left\{
  \begin{array}{lcl}
  \check\rho_{n} \hat\tau_{n} - \check\rho_{n-1} \hat\tau_{n+1} & = & 
  \mybetaB \; \rho_{n}\tau_{n} 
  \\[2mm]
  \check\tau_{n} \hat\sigma_{n} - \check\tau_{n-1} \hat\sigma_{n+1} & = & 
  \mybetaC \; \tau_{n}\sigma_{n} 
  \end{array} 
  \right.
\label{eq-tau-beta}
\end{equation} 
and
\begin{equation}
  \left\{
  \begin{array}{lcl}
  \check\tau_{n-1} \hat\tau_{n} + \check\rho_{n-1} \hat\sigma_{n} & = & 
  \mygammaA \; \tau_{n-1}\tau_{n} 
  \\[2mm]
  \check\tau_{n} \hat\tau_{n} + \check\rho_{n-1} \hat\sigma_{n+1} & = & 
  \mygammaB \; \tau_{n}^{2} 
  \\[2mm]
  \check\tau_{n-1} \hat\tau_{n+1} + \check\rho_{n} \hat\sigma_{n} & = & 
  \mygammaC \; \tau_{n}^{2} 
  \end{array} 
  \right.
\label{eq-tau-gamma}
\end{equation}
where $\mybetaB$, $\mygammaA$, $\mygammaB$ and 
$\mygammaC$ are again some constants. 
It can be shown by direct calculations that Eqs. 
(\ref{eq-tau-alpha})--(\ref{eq-tau-gamma}) imply that $\myB_{n}$, $\myC_{n}$ 
satisfy Eqs. (\ref{system-P}), (\ref{system-N}). 
Indeed, noting that Eqs. (\ref{eq-tau-beta}) and (\ref{eq-tau-gamma}) are nothing but 
\begin{eqnarray} 
  \myB_{n+1} - \myB_{n} & = & 
  \mybetaB \; 
  \displaystyle\frac{ \rho_{n}\tau_{n} }{ \hat\tau_{n}\hat\tau_{n+1} }
  \\[2mm]
  \myC_{n+1} - \myC_{n} & = & 
  \mybetaC \; 
  \displaystyle\frac{ \tau_{n}\sigma_{n} }{ \check\tau_{n-1}\check\tau_{n} }
\end{eqnarray}
and
\begin{eqnarray}
  1 - \myB_{n}\myC_{n} & = & 
  \mygammaA \; 
  \displaystyle\frac{ \tau_{n-1}\tau_{n} }{ \check\tau_{n-1}\hat\tau_{n} }
  \\[2mm]
  1 - \myB_{n}\myC_{n+1} & = & 
  \mygammaB \; 
  \displaystyle\frac{ \tau_{n}^{2} }{ \check\tau_{n}\hat\tau_{n} }
  \\[2mm]
  1 - \myB_{n+1}\myC_{n} & = & 
  \mygammaC \; 
  \displaystyle\frac{ \tau_{n}^{2} }{ \check\tau_{n-1}\hat\tau_{n+1} },
\end{eqnarray}
calculating $f_{n}$,
\begin{equation}
  f_{n} =
  \frac{ \left( \mygammaA \right)^{2} }{ \mygammaB\mygammaC } \; 
  \frac{ \tau_{n-1}\tau_{n+1} }{ \tau_{n}^{2} },
\end{equation}
and substituting the above formulae into (\ref{eq-tau-alpha}) 
and (\ref{eq-tau-bar-alpha}) one can obtain 
\begin{eqnarray}
  i \partial \myB_{n} & = & 
  \Gamma^{B} \; 
  \frac{ 1 - \myB_{n}\myC_{n} }{ 1 - \myB_{n-1}\myC_{n} } 
  \left( \myB_{n} - \myB_{n-1} \right) 
\label{d-myb}
\\[2mm]
  i \partial \myC_{n} & = & 
  \Gamma^{C} \;  
  \frac{ 1 - \myB_{n}\myC_{n} }{ 1 - \myB_{n}\myC_{n+1} } 
  \left( \myC_{n+1} - \myC_{n} \right) 
\end{eqnarray}
and
\begin{eqnarray}
  -i \bar\partial \myB_{n} & = & 
  \bar\Gamma^{B} \; 
  \frac{ 1 - \myB_{n}\myC_{n} }{ 1 - \myB_{n+1}\myC_{n} } 
  \left( \myB_{n+1} - \myB_{n} \right) 
\\[2mm]
  -i \bar\partial \myC_{n} & = & 
  \bar\Gamma^{C} \; 
  \frac{ 1 - \myB_{n}\myC_{n} }{ 1 - \myB_{n}\myC_{n-1} } 
  \left( \myC_{n} - \myC_{n-1} \right) 
\label{bar-d-myc}
\end{eqnarray} 
where
\begin{equation}
  \Gamma^{B} = 
  \frac{ \mygammaB }{ \mygammaA }
  \frac{ \alpha }{ \mybetaB },
  \qquad
  \Gamma^{C} = 
  \frac{ \mygammaB }{ \mygammaA }
  \frac{ \alpha }{ \mybetaC },
  \qquad
  \bar\Gamma^{B} = 
  \frac{ \mygammaC }{ \mygammaA }
  \frac{ \bar\alpha }{ \mybetaB },
  \qquad
  \bar\Gamma^{C} = 
  \frac{ \mygammaC }{ \mygammaA }
  \frac{ \bar\alpha }{ \mybetaC }.
\end{equation}
Thus, to finish solution of our problem one has impose the condition 
\begin{equation}
  \Gamma^{B} = 
  \Gamma^{C} = 
  \zeta,
  \qquad
  \bar\Gamma^{B} = 
  \bar\Gamma^{C} = 
  \bar\zeta = 1 / \zeta.
\end{equation}

In this way  we have reduced 
Eqs. (\ref{system-P}), (\ref{system-N}), and hence Eqs. (\ref{eq-main-matrix}), 
to the set of the bilinear equations (\ref{eq-tau-alpha})--(\ref{eq-tau-gamma}).
An important question that arises now is the question about compatibility of 
this system. We do not present here an explicit proof of the fact that Eqs. 
(\ref{eq-tau-alpha})--(\ref{eq-tau-gamma}) are compatible because (i) we present 
(in the next section) their explicit solutions and (ii) show their relation to a 
well-known nonlinear compatible system -- the Ablowitz-Ladik hierarchy (ALH) 
\cite{AL1975}.
To do the latter let us consider the matrix
\begin{equation}
  \Phi_{n} = 
  \frac{ 1 }{ \tau_{n-1} } 
  \mymatrix{ 
    \hat\tau_{n}     & \check\rho_{n-1} \cr
    - \hat\sigma_{n} & \check\tau_{n-1} 
  }.
\end{equation}
Calculating its determinant,
\begin{equation}
  \det\Phi_{n} = \mygammaA \frac{ \tau_{n} }{ \tau_{n-1} },
\end{equation}
and inverse one can obtain
\begin{equation}
  \Phi_{n+1} = U_{n} \Phi_{n}
\label{alh-sp}
\end{equation}
with
\begin{equation}
  U_{n} = 
  \frac{ 1 }{ \mygammaA } 
  \mymatrix{ 
    \mygammaC     & 
    \mybetaB \displaystyle\frac{ \rho_{n} }{ \tau_{n} } \cr
    \mybetaC \displaystyle\frac{ \sigma_{n} }{ \tau_{n} } &
    \mygammaB 
  }
\label{alh-u}
\end{equation}
and
\begin{equation}
  i\partial\Phi_{n} = V_{n} \Phi_{n}
  \hspace{20mm}
  i\bar\partial\Phi_{n} = \bar{V}_{n} \Phi_{n}
\label{alh-evol}
\end{equation}
with
\begin{equation}
  V_{n}= 
  \mymatrix{ 
    i \partial\ln\displaystyle\frac{ \hat\tau_{n} }{ \tau_{n-1} } 
    + \displaystyle\frac{ \alpha }{ \mygammaA } \;
      \displaystyle\frac{ \rho_{n-1}\hat\sigma_{n} }{ \tau_{n-1}\hat\tau_{n} } 
    &
    \displaystyle\frac{ \alpha }{ \mygammaA } \;
    \displaystyle\frac{ \rho_{n-1} }{ \tau_{n-1} } &
    \cr\cr
    \displaystyle\frac{ \alpha }{ \mygammaA } \;
    \displaystyle\frac{ \sigma_{n} }{ \tau_{n} } 
    &
    i \partial\ln\displaystyle\frac{ \check\tau_{n-1} }{ \tau_{n-1} } 
    - \displaystyle\frac{ \alpha }{ \mygammaA } \;
      \displaystyle\frac{ \check\rho_{n-1} \sigma_{n} }{ \check\tau_{n-1} \tau_{n} } 
  }
\label{alh-v-pos}
\end{equation}
and
\begin{equation}
  \bar{V}_{n}= 
  \mymatrix{ 
    i \bar\partial\ln\displaystyle\frac{ \hat\tau_{n} }{ \tau_{n-1} } 
    - \displaystyle\frac{ \bar\alpha }{ \mygammaA } \;
      \displaystyle\frac{ \rho_{n}\hat\sigma_{n} }{ \tau_{n}\hat\tau_{n} } 
    &
    - \displaystyle\frac{ \bar\alpha }{ \mygammaA } \;
      \displaystyle\frac{ \rho_{n} }{ \tau_{n} } &
    \cr\cr
    - \displaystyle\frac{ \bar\alpha }{ \mygammaA } \;
      \displaystyle\frac{ \sigma_{n-1} }{ \tau_{n-1} } 
    &
    i \bar\partial\ln\displaystyle\frac{ \check\tau_{n-1} }{ \tau_{n-1} } 
    + \displaystyle\frac{ \bar\alpha }{ \mygammaA } \;
      \displaystyle\frac{ \check\rho_{n-1} \sigma_{n-1} }{ \check\tau_{n-1} \tau_{n-1} } 
  }
\label{alh-v-neg}
\end{equation}
Inspecting (\ref{alh-sp})--(\ref{alh-v-neg}) one can conclude, after eliminating the 
unnecessary constants, introducing
\begin{equation}
  q_{n} = \frac{ \sigma_{n} }{ \tau_{n} },
  \qquad
  r_{n} = \frac{ \rho_{n} }{ \tau_{n} } 
\end{equation}
and making some simple gauge transformations, 
that (\ref{alh-sp}) with (\ref{alh-u}) is nothing but the spectral problem of 
the ALH whereas Eqs. (\ref{alh-evol}) with (\ref{alh-v-pos}) and
(\ref{alh-v-neg}) describe its first positive and negative flows.
So, bilinear equations (\ref{eq-tau-alpha})--(\ref{eq-tau-gamma}) belong to 
the ALH. This leads to two important results: (i) they are compatible and 
(ii) we can use already known solutions for the ALH to get solutions of 
our equations.

To expose the inner structure of Eqs. (\ref{eq-tau-alpha})--(\ref{eq-tau-gamma}) 
and to make the following formulae more readable it seems useful to introduce 
instead of the triplet $\rho_{n}$, $\tau_{n}$ and $\sigma_{n}$ an infinite set 
of tau-functions $\tau^{m}_{n}$, 
\begin{equation}
	\rho_{n} = \tau^{-1}_{n}, \qquad 
	\tau_{n} = \tau^{0}_{n} , \qquad
	\sigma_{n} = \tau^{1}_{n}.
\end{equation}
In new terms equations (\ref{eq-tau-alpha})--(\ref{eq-tau-gamma}) become
\begin{eqnarray}
  i D \, \check\tau^{m-1}_{n-1} \cdot \hat\tau^{m}_{n} & = & 
  \alpha \; \tau^{m-1}_{n-1}\tau^{m}_{n} 
\label{bilin-alpha}
\\[2mm]
  -i \bar{D} \, \check\tau^{m-1}_{n-1} \cdot \hat\tau^{m}_{n} & = & 
  \bar\alpha \; \tau^{m}_{n-1}\tau^{m-1}_{n} 
\label{bilin-bar-alpha}
\\[2mm]
  \check\tau^{m-1}_{n} \hat\tau^{m}_{n} - 
  \check\tau^{m-1}_{n-1} \hat\tau^{m}_{n+1} & = & 
  \mybetaB \; \tau^{m-1}_{n}\tau^{m}_{n} 
\label{bilin-beta}
\end{eqnarray} 
for $m=0,1$ and
\begin{eqnarray}
  \check\tau^{m}_{n-1} \hat\tau^{m}_{n} 
  +  
  \check\tau^{m-1}_{n-1} \hat\tau^{m+1}_{n} 
  & = & 
  \mygammaA\; 
  \tau^{m}_{n-1} \tau^{m}_{n} 
\label{bilin-gamma-A}
\\[2mm]
  \check\tau^{m}_{n} \hat\tau^{m}_{n} 
  + 
  \check\tau^{m-1}_{n-1} \hat\tau^{m+1}_{n+1} 
  & = & 
  \mygammaB\; 
  \left( \tau^{m}_{n} \right)^{2}  
\label{bilin-gamma-B}
\\[2mm]
  \check\tau^{m-1}_{n} \hat\tau^{m+1}_{n} 
  + 
  \check\tau^{m}_{n-1} \hat\tau^{m}_{n+1} 
  & = & 
  \mygammaC\; 
  \left( \tau^{m}_{n} \right)^{2}
\label{bilin-gamma-C}
\end{eqnarray}
for $m=0$. These equations are a part of the generalized ALH \cite{V2002} and 
can be solved without imposing restrictions on $m$, for 
$-\infty < m < \infty$.

\section{Dark solitons. \label{sec-solitons}}

\subsection{Dark solitons of the ALH. }

Here we would like to present some basic formulae describing the dark-soliton 
solutions of the ALH that we then use to obtain solutions of our problem.

The dark solitons for the AL equations were obtained in \cite{VK1992} using 
the inverse scattering method. In \cite{V2010} these solutions were derived, 
using purely algebraic method based on the Fay-like identities for the 
determinants of some special matrices.
Here we use notation slightly different from one of \cite{V2010}, which 
makes the following formulae more simple and clear.

The key objects behind the dark-soliton solutions of the ALH are the determinants 
\begin{equation}
  \omega\left( A \right) = \det \left| \mathbb{I} + A \right|
\end{equation}
with matrices $A$ satisfying 
\begin{equation}
  L A - A R = | \,\ell\, \rangle \langle a |.
\label{rank-one-LR}
\end{equation}
Here $\mathbb{I}$ is the $N \times N$ unit matrix, 
$L$ and $R$ are constant diagonal matrices, 
\begin{equation}
  \begin{array}{lcl}
  L & = & \mbox{diag}\left( L_{1}, ..., L_{N} \right), 
  \\
  R & = & \mbox{diag}\left( R_{1}, ..., R_{N} \right),
  \end{array} 
\end{equation}
$| \,\ell\, \rangle$ is a constant $N$-column, 
$| \,\ell\, \rangle = \left( \ell_{1}, ... , \ell_{N} \right)^{T}$,
and $\langle a |$ is a $N$-row depending on the coordinates 
describing the ALH flows: in our case 
$\langle a | = \langle a\left(z, \bar{z}\right) | = 
\left( a_{1}\left(z, \bar{z}\right), ... , a_{N}\left(z, \bar{z}\right) \right)$. 
In what follows we use 'shifted' determinants

\begin{equation}
  \omega_{\zeta} = \mathbb{T}_{\zeta}\,\omega, 
  \qquad
  \omega_{\xi\eta} = \mathbb{T}_{\xi}\mathbb{T}_{\eta}\,\omega
\end{equation}
where
\begin{equation}
  \mathbb{T}_{\zeta}^{l} \, \omega = 
  \omega\left( A \, H_{\zeta}^{l} \right),
  \qquad
  l = \pm 1
\end{equation}
with
\begin{equation}
  H_{\zeta} = 
  \left( L - \zeta \mathbb{I} \right) 
  \left( R - \zeta \mathbb{I} \right)^{-1}.
\end{equation}
An important property of these determinants, that we repeatedly use below, 
is the Fay's identity
\begin{equation} 
  (\xi   - \eta)  \, \omega_{\zeta} \, \omega_{\xi\eta} 
  + 
  (\eta  - \zeta) \, \omega_{\xi}   \, \omega_{\eta\zeta} 
  +  
  (\zeta - \xi)   \, \omega_{\eta}  \, \omega_{\zeta\xi} 
  = 0 
\label{fay}
\end{equation}
which can be proved directly.

Using the limit procedure one can introduce differential operators 
$\partial_{\zeta}$ as
\begin{equation}
  \mathbb{T}_{\zeta}^{-1} \, \mathbb{T}_{\zeta+\delta} \, \omega = 
  \omega + i\delta \, \partial_{\zeta} \omega + 
  O\left( \delta^{2} \right) 
\end{equation}
or
\begin{equation}
  i \partial_{\zeta} A = A X_{\zeta} 
\end{equation}
where
\begin{equation}
  X_{\zeta} = 
  \left( L - R \right) 
  \left( L - \zeta \mathbb{I} \right)^{-1} 
  \left( R - \zeta \mathbb{I} \right)^{-1}.
\end{equation}
One can obtain from (\ref{fay}) many differential Fay's identities of the 
following type:
\begin{equation}
  i (\zeta - \alpha)(\zeta - \beta) \; 
  D_{\zeta} \, \omega_{\alpha} \cdot \omega_{\beta}
  = 
  (\alpha - \beta) 
  \left[
    \left( \mathbb{T}_{\zeta}^{-1} \omega_{\alpha\beta} \right) 
    \left( \mathbb{T}_{\zeta} \omega \right) 
    -
    \omega_{\alpha} \omega_{\beta}
  \right]
\label{fay-diff}
\end{equation}
where
\begin{equation}
  D_{\zeta} \, \omega_{\alpha} \cdot \omega_{\beta}
  = 
  \left( \partial_{\zeta} \omega_{\alpha} \right) \omega_{\beta} 
  - 
  \omega_{\alpha} \left( \partial_{\zeta} \omega_{\beta} \right).
\end{equation}

The matrices $L$ and $R$ used in the ALH context are not independent: they are related by 
\begin{equation}
  ( L - \kappa \mathbb{I} )( R - \kappa \mathbb{I} ) = - \rho^{2} \, \mathbb{I}. 
\label{rel-LR}
\end{equation}
with constant parameters $\kappa$ and $\rho$.
Relations of this kind play crucial role in the construction of 
dark solitons for the ALH, so it seems useful introduce the notion of 'duality': 
two complex numbers   $\xi$ and $\xi^{*}$ are said to be dual if 
\begin{equation}
  ( \zeta - \kappa )( \zeta^{*} - \kappa ) = - \rho^{2}. 
\label{duality}
\end{equation}
which leads to an alternative definition 
\begin{equation}
  H_{\zeta} H_{\zeta^{*}} = H_{\kappa}, 
  \qquad
  \kappa = \mbox{const}
\end{equation}
Many Fay-like formulae can be simplified when rewritten in terms of dual 
numbers. In particular, Eq. (\ref{fay-diff}) leads to 
\begin{equation}
  i D_{\zeta} \, \omega_{\alpha} \cdot \omega_{\alpha^{*}}
  = 
  \frac{ \alpha - \alpha^{*} }
       { \left( \zeta - \alpha \right) \left( \zeta - \alpha^{*} \right) } 
  \left[
    \left( \mathbb{T}_{\zeta}^{-1} \omega_{\kappa} \right) 
    \left( \mathbb{T}_{\zeta} \omega \right) 
    -
    \omega_{\alpha} \omega_{\alpha^{*}}
  \right]
\end{equation}

Given some fixed number $\mu$ and its dual, which will be denoted by $\nu$,  
$\nu = \mu^{*}$, one can construct an doubly infinite set of 
matrices/determinants  
\begin{equation}
  \omega^{m}_{n} = \omega\left( A^{m}_{n} \right)
\end{equation}
where
\begin{equation}
  A^{m}_{n} = A H_{\mu}^{m} H_{\nu}^{n} 
\label{matrix-A-mn}
\end{equation}
and derive from (\ref{fay}) a lot of lattice Fay's identities the most 
important of which is 
\begin{equation}
  \left( \omega^{m}_{n} \right)^{2} = 
  \rho_{\mu}^{2} \omega^{m-1}_{n} \, \omega^{m+1}_{n} 
  + 
  \rho_{\nu}^{2} \omega^{m}_{n-1} \, \omega^{m}_{n+1} 
\end{equation}
with
\begin{equation}
  \rho_{\mu} = \sqrt{ \frac{ \mu - \kappa } { \mu - \nu } }, 
  \qquad 
  \rho_{\nu} = \sqrt{ \frac{ \kappa - \nu }{ \mu - \nu } }.
\end{equation}

In order to ensure the involution $\overline{\tau^{m}_{n}} = \tau^{-m}_{n}$ 
(where overline stands for the complex conjugation) 
which appears in physical applications of the Ablowitz-Ladik model one has 
to restrict himself to the case of real $\kappa$, $\mu$ and $\nu$,
\begin{equation}
  \nu < \kappa < \mu
\end{equation}
(which leads to $\Im\rho_{\mu} = \Im\rho_{\nu} = 0$)
and to choose the matrices $L$ to be of the form 
\begin{equation}
  L = \mu + \sqrt{ \mu - \nu } \sqrt{ \mu - \kappa } \, E,
  \qquad
  E = \mbox{diag}\left( e^{i\psi_{j}} \right) 
\label{parametrization-L}
\end{equation}
with real angles $\psi_{j}$ (compare with the parametrization (2.27) of the 
eigenvalues of the scattering problem for the ALH used in \cite{VK1992}). 

Calculating $R$ from (\ref{parametrization-L}) and (\ref{rel-LR}) one 
can verify that in this case
\begin{equation}
  \overline{H_{\mu}} = H_{\mu}^{-1},
  \qquad
  \overline{H_{\nu}} = H_{\nu} 
\end{equation}
and 
\begin{equation}
  \overline{A^{m}_{n}} = A^{-m}_{n}
\end{equation}
provided
\begin{equation}
  \overline{A} = A.
\end{equation}
The last condition can be met by choosing properly the constants 
$\ell_{j}$ and $a_{j}$ in (\ref{rank-one-LR}). 
By straightforward algebra one can get 
\begin{equation}
  L_{j} - R_{k} = 
  ( \mu - \nu ) \;
  \frac{ e^{i\left(\psi_{j}+\psi_{k}\right)} 
         + \rho_{\mu} \left( e^{i\psi_{j}} + e^{i\psi_{k}} \right) 
         + 1 }
       { 1 + \rho_{\mu}^{-1} e^{i\psi_{k}} }  
\end{equation}
and rewrite the matrix $A$, after eliminating excessive constants, as
\begin{equation}
  A(z,\bar{z})_{jk} = 
  D_{jk} c_{k}(z,\bar{z})
\end{equation}
where
\begin{equation}
  D_{jk} = 
  \left[ 
    \cos\left( \frac{\psi_{j} + \psi_{k}}{ 2 } \right) + 
    \rho_{\mu} \cos\left( \frac{\psi_{j} - \psi_{k}}{ 2 } \right) 
  \right]^{-1}
\label{def-Djk}
\end{equation}
and $c_{k}(z,\bar{z})$ are some real functions.

\subsection{Dark soliton solutions of (\ref{bilin-alpha})--(\ref{bilin-gamma-C}). }

After we have established the relation of our model with the ALH and knowing 
the structure of the ALH dark solitons, we can reformulate the \textit{ansatz} 
we use as follows: all tau-functions are related by the $\mathbb{T}$-shifts 
($\mathbb{T}_{\nu}$, $\mathbb{T}_{\mu}$ and $\mathbb{T}_{\xi^{*}}$ for some 
given $\xi$). The sequence $\rho \to \tau \to \sigma$ is generated by 
$\mathbb{T}_{\mu}$, the sequence $\check\tau \to \tau \to \hat\tau$ is generated by 
$\mathbb{T}_{\xi^{*}}$, 
as is depicted in Fig. 1, 
 while the nodes $n$ and $n+1$ are related by 
$\mathbb{T}_{\nu}$. Thus we can say that our tau-functions occupy sites of a 
three-dimensional lattice. However in what follows we do not use the 
three-indices and adhere to the $\tau^{m}_{n}$ notation.

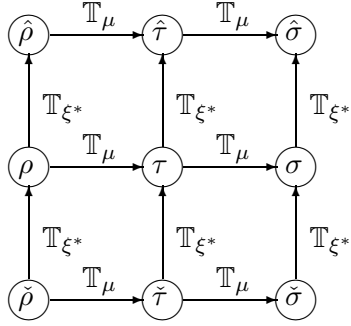
\begin{figure}
\begin{picture}(120,120)(-150,-5)
\put(0,2){$\check\rho$} \put(4,4){\circle{15}}
\put(50,2){$\check\tau$} \put(54,4){\circle{15}}
\put(100,2){$\check\sigma$} \put(104,4){\circle{15}}
\put(12,4){\vector(1,0){35}}
\put(24,9){$\mathbb{T}_{\mu}$}
\put(62,4){\vector(1,0){35}}
\put(74,9){$\mathbb{T}_{\mu}$}
\put(4,12){\vector(0,1){35}}
\put(9,24){$\mathbb{T}_{\xi^{*}}$}
\put(54,12){\vector(0,1){35}}
\put(59,24){$\mathbb{T}_{\xi^{*}}$}
\put(104,12){\vector(0,1){35}}
\put(109,24){$\mathbb{T}_{\xi^{*}}$}
\put(0,52){$\rho$} \put(4,54){\circle{15}}
\put(50,52){$\tau$} \put(54,54){\circle{15}}
\put(100,52){$\sigma$} \put(104,54){\circle{15}}
\put(12,54){\vector(1,0){35}}
\put(24,59){$\mathbb{T}_{\mu}$}
\put(62,54){\vector(1,0){35}}
\put(74,59){$\mathbb{T}_{\mu}$}
\put(4,62){\vector(0,1){35}}
\put(9,74){$\mathbb{T}_{\xi^{*}}$}
\put(54,62){\vector(0,1){35}}
\put(59,74){$\mathbb{T}_{\xi^{*}}$}
\put(104,62){\vector(0,1){35}}
\put(109,74){$\mathbb{T}_{\xi^{*}}$}
\put(0,102){$\hat\rho$} \put(4,104){\circle{15}}
\put(50,102){$\hat\tau$} \put(54,104){\circle{15}}
\put(100,102){$\hat\sigma$} \put(104,104){\circle{15}}
\put(12,104){\vector(1,0){35}}
\put(24,109){$\mathbb{T}_{\mu}$}
\put(62,104){\vector(1,0){35}}
\put(74,109){$\mathbb{T}_{\mu}$}
\end{picture}
\caption{$\mathbb{T}_{\mu}$-$\mathbb{T}_{\xi^{*}}$ lattice of tau-functions 
(for $n$=constant). }
\end{figure}

To find solutions of (\ref{bilin-alpha})--(\ref{bilin-gamma-C}) we look for our 
tau-functions in the form
\begin{equation}
  \tau^{m}_{n} = 
  \rho_{\mu}^{m^{2}} \rho_{\nu}^{n^{2}}
  u^{m} v^{n} 
  \omega^{m}_{n}
\label{def-tau}
\end{equation}
with similar formulae for $\check\tau^{m}_{n}$ and $\hat\tau^{m}_{n}$
\begin{eqnarray}
  \check\tau^{m}_{n} & = & 
  \rho_{\mu}^{m^{2}} \rho_{\nu}^{n^{2}}
  \check{u}^{m} \check{v}^{n} 
  \check\omega^{m}_{n}
\\
  \hat\tau^{m}_{n} & = & 
  \rho_{\mu}^{m^{2}} \rho_{\nu}^{n^{2}}
  \hat{u}^{m} \hat{v}^{n} 
  \hat\omega^{m}_{n}
\end{eqnarray}
where $\check\omega^{m}_{n}$ and $\hat\omega^{m}_{n}$ are related to 
$\omega^{m}_{n}$ by means of $\mathbb{T}_{\xi^{*}}^{\pm 1}$ shifts:
\begin{equation}
  \check\omega^{m}_{n} 
  = 
  \mathbb{T}_{\xi^{*}} \omega^{m}_{n} ,
  \qquad
  \hat\omega^{m}_{n} 
  =  
  \mathbb{T}_{\xi^{*}}^{-1} \omega^{m}_{n}
\end{equation}
or
\begin{equation}
  \check\omega^{m}_{n} 
  = 
  \mathbb{T}_{\xi}^{-1} \omega^{m+1}_{n+1},
  \qquad
  \hat\omega^{m}_{n} 
  = 
  \mathbb{T}_{\xi} \omega^{m-1}_{n-1}. 
\end{equation}
Taking $\xi$, $\eta$, $\zeta$ in (\ref{fay}) being equal to different triples 
from $\left\{ \mu, \nu, \kappa, \xi, \xi^{*} \right\}$ one can conclude that 
restrictions 
\begin{equation}
  \left( \rho_{\mu} \frac{ \hat{u} }{ u } \right)^{2} =
  - \frac{ \xi - \mu }{ \xi^{*} - \mu },
  \hspace{10mm}
  \left( \rho_{\nu} \frac{ \hat{v} }{ v } \right)^{2} =
  \frac{ \xi - \nu }{ \xi^{*} - \nu },
  \hspace{10mm}
  \left( 
    \rho_{\mu}\rho_{\nu} 
    \frac{ \hat{u} }{ u } 
    \frac{ \hat{v} }{ v } 
  \right)^{2} =
  - \frac{ \xi - \kappa }{ \xi^{*} - \kappa }
\end{equation}
lead to (\ref{bilin-beta})--(\ref{bilin-gamma-C}) with
\begin{eqnarray}
  \mybetaB{} 
  & = & 
  \displaystyle 
  \frac{ \xi^{*} - \xi }{ \xi^{*} - \nu } \; 
  \frac{ \hat{u} }{ u } 
\\[4mm]
  \mygammaA{} 
  & = & 
  \displaystyle 
  \frac{ \xi^{*} - \xi }{ \xi^{*} - \mu } \; 
  \frac{ \hat{v} }{ v } 
\\[4mm]
  \mygammaB{} 
  & = & 
  \displaystyle 
  \frac{ \xi^{*} - \xi }{ \xi^{*} - \kappa } 
\\[4mm]
  \mygammaC{} 
  & = & 
  \displaystyle 
  \frac{ \left( \xi^{*} - \xi \right)\left( \mu - \nu \right) }
       { \left( \xi^{*} - \mu \right)\left( \xi^{*} - \nu \right) } 
\end{eqnarray}
Considering the dependence on $z$ and $\bar{z}$ it can be shown that to meet 
(\ref{bilin-alpha}) and (\ref{bilin-bar-alpha}) one has to take 
\begin{equation}
  i\partial A = A X,
  \qquad
  -i\bar\partial A = A \widetilde{X}
\end{equation}
where
\begin{equation}
  X = \mathop{\mbox{const}} \cdot X_{\kappa},
  \qquad
  \widetilde{X} = \mathop{\mbox{const}} \cdot X_{\nu}
\end{equation}
We will write the explicit value of the corresponding constants later, after 
discussing the involution (complex conjugation) and reality requirement.

\subsection{Complex conjugation and parametrization. }

Recalling the definitions of our tau-functions we can present $\myB_{n}$ and 
$\myC_{n}$ as 
\begin{equation}
  \myB_{n} = \myB^{0}_{n}, \qquad \myC_{n} = \myC^{0}_{n}
\end{equation}
where
\begin{eqnarray}
  \myB_{n}^{m} & = & 
  \frac{ 1 }{ uv }
  \left( \frac{ \mu - \xi^{*} }{ \xi - \mu } \right)^{m - \frac{1}{2}}
  \left( \frac{ \xi^{*} - \nu }{ \xi - \nu } \right)^{n - \frac{1}{2}}
  \frac{ \check\omega^{m-1}_{n-1} }{ \hat\omega^{m}_{n} }
\label{my-b-mn}
\\[4mm]
  \myC_{n}^{m} & = & 
  - uv 
  \left( \frac{ \xi - \mu }{ \mu - \xi^{*} } \right)^{m + \frac{1}{2}}
  \left( \frac{ \xi - \nu }{ \xi^{*} - \nu } \right)^{n - \frac{1}{2}}
  \frac{ \hat\omega^{m+1}_{n} }{ \check\omega^{m}_{n-1} }
\label{my-c-mn}
\end{eqnarray}
The involution that ensures reality of $\vec{\sigma}_{n}$ and that is 
consistent with the ALH is 
\begin{equation}
  \myB_{n}^{m} = - \overline{ \myC_{n}^{-m} }
\end{equation}
where overline stands for complex conjugation. A simple analysis of 
(\ref{my-b-mn}), (\ref{my-c-mn}) and (\ref{matrix-A-mn}) leads to the 
restrictions 
\begin{equation}
  \mbox{Im} \, \frac{ \xi^{*} - \mu }{ \xi - \mu } = 0,
  \qquad
  \left| \frac{ \xi^{*} - \nu }{ \xi - \nu } \right| = 1
\label{involution-xi}
\end{equation}
and
\begin{equation}
  H_{\xi} \overline{H_{\xi}} = H_{\nu}  
\label{involution-H-xi}
\end{equation}
that should be added to the restriction $\overline{A^{m}_{n}} = A^{-m}_{n}$ 
discussed above.

The family of $\xi$ and $\xi^{*}$ that satisfy (\ref{duality}), 
(\ref{involution-H-xi}) and (\ref{involution-xi}) can be parametrized as 
\begin{equation}
  \begin{array}{lclcl}
  \xi & = & 
  \nu + a e^{i\left( \phi + \theta \right)} 
  \\[2mm]
  \xi^{*} & = & 
  \nu + a e^{i\left( \phi - \theta \right)} 
  \end{array}
\label{xi-nu}
\end{equation}
where
\begin{equation}
  a = \sqrt{ \mu - \nu } \sqrt{ \kappa - \nu } = \frac{\rho}{\rho_{\mu}}
\end{equation}
and the angles $\phi$ and $\theta$ are related by 
\begin{equation}
  \cos\phi = \rho_{\nu} \cos\theta
\end{equation}
Using another parametrization of the matrices $L$ and $R$, stemming from 
(\ref{parametrization-L}) and (\ref{rel-LR}),
\begin{equation}
  \begin{array}{lcl}
  L_{j} & = & 
  \nu + a \exp\left( \chi_{j} + i \phi_{j} \right)
  \\
  R_{j} & = & 
  \nu + a \exp\left( - \chi_{j} + i \phi_{j} \right)
  \end{array}
\label{LRj-nu}
\end{equation}
where the quantities $\chi_{j}$ and $\phi_{j}$ are defined by 
\begin{equation}
  \exp\left( \chi_{j} + i \theta_{j} \right) = 
  \frac{ 1 }{ \rho_{\nu} } \left( \rho_{\mu} + e^{i\psi_{j}} \right) 
\end{equation}
one can obtain 
\begin{equation}
  H_{\xi} = 
  - \mbox{diag} \left( \; e^{\chi_{j} - i\gamma^{+}_{j}} \; \right), 
\qquad
  H_{\xi^{*}}  = 
  - \mbox{diag} \left( \; e^{\chi_{j} - i\gamma^{-}_{j}} \; \right) 
\end{equation}
with
\begin{equation}
  \gamma^{\pm}_{j} = 
  \phi - \phi_{j} \pm \theta 
  - 2 \arg\left[ 
      e^{\chi_{j}} - e^{ i\left( \phi - \phi_{j} \pm \theta \right) }
    \right].
\label{def-gamma-pm}
\end{equation}

\subsection{Dark solitons of the Toda-Heisenberg chain. }

Now we have all necessary to write dark soliton solutions of the Toda-Heisenberg 
chain. Their structure is given by (\ref{my-b-mn}) and (\ref{my-c-mn}). 
The dependence on $z$ and $\bar{z}$ enters through the matrices $A$ and the 
factor $uv$,
\begin{equation}
  i\partial A = A X,
  \qquad
  -i\bar\partial A = A \widetilde{X}
\end{equation}
where
\begin{equation}
  X = 
  - \frac{ \lambda\rho^{2} }{ \mu - \nu } X_{\kappa}, 
  \qquad
  \widetilde{X} = 
  \lambda^{-1}(\kappa - \nu) \; X_{\nu} 
\end{equation}
or, explicitly,  
\begin{equation}
  X =  
  \frac{ \lambda }{ \mu - \nu }
  \left( L - R \right),
  \qquad
  \widetilde{X} = 
  \frac{ \lambda^{-1} }{ \mu - \nu }
  \overline{\left( L - R \right)}.
\end{equation}
This dependence of $A$, 
as follows from the differential Fay's identity (\ref{fay-diff}), leads to 
\begin{eqnarray}
  i D \, \omega_{\alpha^{*}} \cdot \omega_{\alpha} & = & 
  \frac{ \lambda \left( \alpha^{*} - \alpha \right) }{ \mu - \nu }
  \left[ \omega \, \omega_{\kappa} - \omega_{\alpha^{*}} \omega_{\alpha} \right] 
  \\[4mm]
  i \bar{D} \, \omega_{\alpha^{*}} \cdot \omega_{\alpha} & = & 
  \frac{ \lambda^{-1} \overline{\left( \alpha^{*} - \alpha \right)} }
       { \mu - \nu }
  \left[ \omega_{\mu} \omega_{\nu} - \omega_{\alpha^{*}} \omega_{\alpha} \right] 
\end{eqnarray}
and hence to
\begin{eqnarray}
  i D \, \check\omega^{m-1}_{n-1} \cdot \hat\omega^{m}_{n} & = & 
  \frac{ \lambda \left( \xi^{*} - \xi \right) }{ \mu - \nu } 
  \left[ \omega^{m-1}_{n-1} \, \omega^{m}_{n} - 
         \check\omega^{m-1}_{n-1} \, \hat\omega^{m}_{n} \right] 
  \\
  i D \, \hat\omega^{m+1}_{n} \cdot \check\omega^{m}_{n-1} & = & 
  \frac{ \lambda \left( \xi^{*} - \xi \right) }{ \mu - \nu } 
  \left[ \check\omega^{m}_{n-1} \, \hat\omega^{m+1}_{n} - 
         \omega^{m}_{n-1} \, \omega^{m+1}_{n} \right] 
\end{eqnarray}
and
\begin{eqnarray}
  i \bar{D} \, \check\omega^{m-1}_{n-1} \cdot \hat\omega^{m}_{n} & = & 
  \frac{ \lambda^{-1} \overline{\left( \xi^{*} - \xi \right)} }
       { \mu - \nu }
  \left[ \omega^{m-1}_{n} \, \omega^{m}_{n-1} - 
         \check\omega^{m-1}_{n-1} \, \hat\omega^{m}_{n} \right] 
  \\
  i \bar{D} \, \hat\omega^{m+1}_{n} \cdot \check\omega^{m}_{n-1} & = & 
  \frac{ \lambda^{-1} \overline{\left( \xi^{*} - \xi \right)} }
       { \mu - \nu }
  \left[ \check\omega^{m}_{n-1} \, \hat\omega^{m+1}_{n} - 
         \omega^{m}_{n} \, \omega^{m+1}_{n-1} \right]. 
\end{eqnarray}
The extra terms in the right-hand sides of the above formulae can be eliminated 
by taking
\begin{equation}
  uv = \mbox{const} \times e^{i\varphi}
\end{equation}
with
\begin{equation}
  \varphi = 
  \lambda \;
  \displaystyle 
  \frac{ \xi^{*} - \xi }{ \mu - \nu } \; z
  + 
  \lambda^{-1} \;
  \displaystyle 
  \frac{ \overline{ \left(\xi^{*} - \xi\right)} }{ \mu - \nu } \; \bar{z}, 
\end{equation}
It is easy to verify that all reality conditions are met provided 
\begin{equation}
  | \lambda | = 1.
\end{equation}
Now one can check that tau-functions $\tau^{m}_{n}$ satisfy Eqs. 
(\ref{bilin-alpha}) and (\ref{bilin-bar-alpha}) with 
\begin{equation}
  \alpha 
  =  
  \lambda \;
  \frac{ \xi^{*} - \xi }{ \mu - \nu } \;
  \frac{ \hat{u}\hat{v} }{ uv }, 
	\qquad
  \bar\alpha 
  = 
  - \lambda^{-1} 
  \frac{ \overline{ \left(\xi^{*} - \xi\right)} }{ \mu - \nu } \;
  \frac{ \hat{u}\hat{v} }{ uv } 
\end{equation}
Gathering all constants, one comes to the conclusion that for $m=0$ the 
quantities $\myB_{n}$, $\myC_{n}$ solve Eqs. (\ref{d-myb})--(\ref{bar-d-myc})
with 
\begin{equation}
  \Gamma^{B} = \Gamma^{C}
  = 
  - \lambda \;
  e^{ 2i \arg\left(\xi^{*} - \xi\right) }
\end{equation}
and
\begin{equation}
  \bar\Gamma^{B} = \bar\Gamma^{C}
  = 
  - \lambda^{-1} \;
  e^{ -2i \arg\left(\xi^{*} - \xi\right) }
\end{equation}
Finally, calculating from (\ref{xi-nu}) and (\ref{LRj-nu}) the coefficients 
that describe the $z$-, $\bar{z}$-dependence,
\begin{equation}
  \frac{ L_{j} - R_{j} }{ \mu - \nu } 
  = 
  2 \rho_{\mu} \, \cos\theta_{j} \; e^{ i\phi_{j} } 
\end{equation}
and 
\begin{equation}
  \frac{ \xi^{*} - \xi }{ \mu - \nu } 
  = 
  - 2 i \rho_{\nu} \, \sin\theta \; e^{ i\phi }, 
\end{equation}
one can present $\myB_{n}$ and $\myC_{n}$ as follows:
\begin{equation}
  \myB_{n}(x,y) = 
  B_{*} 
  \exp\left\{ - 2in\theta - i \varphi(x,y) \right\} \; 
  \frac{ \Delta^{+}_{n}(x,y) }{ \Delta^{-}_{n}(x,y) } 
\end{equation}
and 
\begin{equation}
  \myC_{n}(x,y) = - \overline{\myB_{n}(x,y)}. 
\end{equation}
Here
\begin{equation}
  B_{*} = 
  \frac{ 1 }{ \rho_{\mu} } 
  \left| 1 - \rho_{\nu} \exp\left( i \psi \right) \right|,
\end{equation}
\begin{equation}
  \Delta^{\pm}_{n}(x,y)  = 
  \det\left|
    \delta_{jk} - 
    D_{jk} \exp\left\{ 2n\chi_{k} + a_{k}(x,y) + i\gamma^{\pm}_{k} \right\}
  \right|,
\end{equation}
with the coefficients $D_{jk}$ and $\gamma^{\pm}_{k}$ being defined in (\ref{def-Djk}), 
(\ref{def-gamma-pm}). 
The phase $\varphi$ and the functions $a_{k}$ are given by
\begin{equation}
  \varphi(x,y) = 
  4 \rho_{\nu} \, \sin\theta 
  \left( x \sin\alpha_{0} + y \cos\alpha_{0} \right) 
  + \varphi^{(0)}
\end{equation}
with 
\begin{equation}
  \alpha_{0} = \phi + \arg\lambda 
\end{equation}
and
\begin{equation}
  a_{k}(x,y) = 
  4 \rho_{\mu}  \, \cos\theta_{k} 
  \left( x\sin\alpha_{k} + y\cos\alpha_{k}\right)  
  + a^{(0)}_{k}
\end{equation}
with
\begin{equation}
  \alpha_{k} = \phi_{k} + \arg\lambda 
\end{equation}
where $\varphi^{(0)}$ and $a^{(0)}_{k}$ are arbitrary real constants.

These formulae, together with the vector-matrix correspondence (\ref{vmc}), 
lead to the dark soliton solutions of the Toda-Heisenberg chain:
\begin{equation}
  \vec{\sigma}_{n} = 
  \frac{ 1 }{ 1 + \left| B_{n} \right|^{2} }
  \begin{pmatrix}
    \phantom{-} 2\Re B_{n} \cr
    -2\Im B_{n} \cr
    1 - \left| B_{n} \right|^{2} 
  \end{pmatrix}.
\end{equation}

\section{Conclusion.}                                                         %

We have studied the (2+1)-dimensional system that was reduced to the 
integrable Ablowitz-Ladik equations. Using this reduction we have derived its 
soliton solutions. It is clear that using this approach one can also derive a 
wide range of other solutions starting from the ones already known for the 
ALH. Thus Eqs. (\ref{eq-htc-vector}), (\ref{def-f-nn}) possess a set of 
solutions that are typical for the integrable systems (solitons, 
algebro-geometric solutions \textit{etc}). At the same time it is not clear 
whether this model is integrable or we deal with another example of soliton 
equation that is not integrable (see, \textit{e.g.}, \cite{PV2010}), 
which can occur in multidimensions, contrary (as is presumed) to the 
(1+1)-dimensional case. However, this very interesting question, as well as 
other related  questions (such as, \textit{e.g.}, the Painlev\'e test, 
the symmetry analysis), is out of the scope of this paper and may constitute 
the subject of the subsequent studies.

\appendix
\section{}

Rewriting the \textit{ansatz} (\ref{system-ansatz-P}), (\ref{system-ansatz-N}) 
in the form 

\begin{equation}
  \left\{
  \begin{array}{lcl}
  i  \partial \, \myB_{n} & = & 
  \zeta_{n-1} Y_{n-1} 
  \left( 1 - \myB_{n}\myC_{n} \right) 
  \left( \myB_{n} - \myB_{n-1} \right) 
  \\
  i \partial \, \myC_{n} & = & 
  \zeta_{n} Y_{n} 
  \left( 1 - \myB_{n}\myC_{n} \right) 
  \left( \myC_{n+1} - \myC_{n} \right) 
  \end{array} 
  \right.
\label{proof-ansatz-P}  
\end{equation}
and
\begin{equation}
  \left\{
  \begin{array}{lcl}
  -i \bar\partial \, \myB_{n} & = & 
  \bar\zeta_{n}\bar{Y}_{n} 
  \left( 1 - \myB_{n}\myC_{n} \right) 
  \left( \myB_{n+1} - \myB_{n} \right) 
  \\
  -i \bar\partial \, \myC_{n} & = & 
  \bar\zeta_{n-1}\bar{Y}_{n-1} 
  \left( 1 - \myB_{n}\myC_{n} \right) 
  \left( \myC_{n} - \myC_{n-1} \right) 
  \end{array} 
  \right.
\label{proof-ansatz-N}  
\end{equation}
one can calculate $\partial\bar\partial \, \myB_{n}$ as
\begin{equation}
  \partial\bar\partial \, \myB_{n} = 
  i\partial \left( -i \bar\partial \, \myB_{n} \right) =
  i \partial 
  \, 
  \bar\zeta_{n}\bar{Y}_{n} 
  \left( 1 - \myB_{n}\myC_{n} \right) 
  \left( \myB_{n+1} - \myB_{n} \right) 
\end{equation}
which leads, after repeated usage of (\ref{proof-ansatz-P}) to 
\begin{equation}
  \myA_{n} \mathcal{L}^{\myB}_{n} 
  = 
  \left( i \partial\bar\zeta_{n} + \zeta_{n} \bar\zeta_{n} \right) \; 
  \bar{Y}_{n} \left( \myB_{n+1} - \myB_{n} \right)
  - 
  \zeta_{n-1} \bar\zeta_{n} \; 
  Y_{n-1} \left( \myB_{n} - \myB_{n-1} \right).
\end{equation}
Interchanging the $\partial$ and $\bar\partial$ derivatives, 
$  \partial\bar\partial \, \myB_{n} = 
  -i \bar\partial \left( i\partial \, \myB_{n} \right)
$, 
one can obtain
\begin{equation}
  \myA_{n} \mathcal{L}^{\myB}_{n} 
   = 
  \zeta_{n-1} \bar\zeta_{n} \; 
  \bar{Y}_{n} \left( \myB_{n+1} - \myB_{n} \right)
  - 
  \left( i \bar\partial\zeta_{n-1} + \zeta_{n-1} \bar\zeta_{n-1} \right) \; 
  Y_{n-1} \left( \myB_{n} - \myB_{n-1} \right).
\end{equation}
Similar calculations for $\partial\bar\partial \, \myC_{n}$ lead to 
\begin{eqnarray}
  \myA_{n} \mathcal{L}^{\myC}_{n} 
  & = & 
  \left( - i \bar\partial\zeta_{n} + \zeta_{n} \bar\zeta_{n} \right) \; 
  Y_{n} \left( \myC_{n+1} - \myC_{n} \right)
  - 
  \zeta_{n} \bar\zeta_{n-1} \; 
  \bar{Y}_{n-1} \left( \myC_{n} - \myC_{n-1} \right)
\\[2mm]
  & = & 
  \zeta_{n} \bar\zeta_{n-1} \; 
  Y_{n} \left( \myC_{n+1} - \myC_{n} \right)
  + 
  \left( i \partial\bar\zeta_{n-1} - \zeta_{n-1} \bar\zeta_{n-1} \right) \; 
  \bar{Y}_{n-1} \left( \myC_{n} - \myC_{n-1} \right)
\end{eqnarray}
Comparing the right-hand sides of the above equations with each other and 
with the right-hand sides of Eqs. (\ref{syst-BC}) one can conclude that 
conditions
\begin{eqnarray}
&&  \zeta_{n \pm 1} = \zeta_{n}, \\
&&  \bar\zeta_{n \pm 1} = \bar\zeta_{n}, \\
&&
  \partial\zeta_{n} = \bar\partial\zeta_{n} = 
  \partial\bar\zeta_{n} = \bar\partial\bar\zeta_{n} = 0
\end{eqnarray}
and
\begin{equation}
  \zeta_{n} \bar\zeta_{n} = 1
\end{equation}
validate the \textit{ansatz} (\ref{system-P}), (\ref{system-N}).


\end{document}